\documentclass[letterpaper, 10 pt, conference]{ieeeconf}  

\IEEEoverridecommandlockouts                              

\usepackage{xcolor}

\usepackage{bm}

\usepackage{amsthm}
\usepackage{amsmath}
\usepackage[english]{babel}
\usepackage{graphicx} 
\usepackage{caption}
\usepackage[colorlinks]{hyperref}
\usepackage{amsfonts}
\usepackage{subcaption}
\newtheorem{definition}{Definition}
\newtheorem{theorem}{Theorem}

\usepackage{eurosym}
\usepackage{dsfont}
\usepackage[nocompress, space]{cite}
\usepackage{textcomp}
\usepackage[displaymath,mathlines]{lineno}

\allowdisplaybreaks

\usepackage{ifthen}

\newcommand{\Rr}{{\mathbb{R}}}

\newcommand{\Aa}{{\mathcal{A}}}

\newtheoremstyle{thmlemcorr}{10pt}{10pt}{\itshape}{}{\bfseries}{.}{10pt}{{\thmname{#1}\thmnumber{
                        #2}\thmnote{ (#3)}}}
\newtheoremstyle{thmlemcorr*}{10pt}{10pt}{\itshape}{}{\bfseries}{.}\newline{{\thmname{#1}\thmnumber{
                        #2}\thmnote{ (#3)}}}
\newtheoremstyle{defi}{10pt}{10pt}{\itshape}{}{\bfseries}{.}{10pt}{{\thmname{#1}\thmnumber{
                        #2}\thmnote{ (#3)}}}
\newtheoremstyle{remexample}{10pt}{10pt}{}{}{\bfseries}{.}{10pt}{{\thmname{#1}\thmnumber{
                        #2}\thmnote{ (#3)}}}
\newtheoremstyle{ass}{10pt}{10pt}{}{}{\bfseries}{.}{10pt}{{\thmname{#1}\thmnumber{
                        A#2}\thmnote{ (#3)}}}

\theoremstyle{thmlemcorr*}
\newtheorem{theorem*}{Theorem}
\newtheorem{lemma*}[theorem]{Lemma}
\newtheorem{corollary*}[theorem]{Corollary}
\newtheorem{proposition*}[theorem]{Proposition}
\newtheorem{problem*}[theorem]{Problem}
\newtheorem{conjecture*}[theorem]{Conjecture}

\newtheorem{problem}{Problem}

\theoremstyle{remexample}
\newtheorem{remark}[theorem]{Remark}

\title{\LARGE \bf
Hessian Riemannian Flow For Multi-Population Wardrop Equilibrium
}

\author{Tigran Bakaryan, Christoph Aoun, Ricardo de Lima Ribeiro,
        Naira Hovakimyan, and Diogo  Gomes
\thanks{The Work of TB was supported by the Higher Education and Science Committee of MESCS RA, Research Project, 24IRF-1A001. The work of CA and NH was supported by National Aeronautics and Space Administration (NASA) University Leadership Initiative (ULI) grant 80NSSC17M0051.}
\thanks{Tigran Bakaryan is with the Center for Scientific Innovation and Education, Institute of Mathematics NAS RA, Yerevan, Armenia.
        {\tt\small tigran.bakaryan@instmath.sci.am}}
\thanks{Christoph Aoun is with the Department of Aerospace Engineering, University of Illinois at Urbana-Champaign, 
        Urbana, IL 61801, USA.  
        {\tt\small caoun2 @illinois.edu}}
\thanks{ Ricardo de Lima Ribeiro is with Applied Mathematics and Computational Sciences, King Abdullah University of Science and Technology, 23955 Thuwal, Saudi Arabia. 
        {\tt\small Arabia; ricardo.ribeiro@kaust.edu.sa}}
\thanks{Naira Hovakimyan is with the Department of Mechanical Science and Engineering, University of Illinois at Urbana-Champaign, 
        Urbana, IL 61801, USA.  
        {\tt\small \{nhovakim\} @illinois.edu}}
\thanks{Diogo  Gomes Applied is with the Mathematics and Computational Sciences, King Abdullah University of Science and
Technology, 23955 Thuwal, Saudi Arabia.
        {\tt\small diogo.gomes@kaust.edu.sa}}
}

\begin{document}
\maketitle
\begin{abstract}
In this paper, we address the problem of optimizing flows on generalized graphs that feature multiple entry points and multiple populations, each with varying cost structures. We tackle this problem by considering the multi-population Wardrop equilibrium, defined through variational inequalities.
We rigorously analyze the existence and uniqueness of the Wardrop equilibrium. Furthermore, we introduce an efficient numerical method to find the solution. In particular, we reformulate the equilibrium problem as a distributed optimization problem over subgraphs and introduce a novel Hessian Riemannian flow method—a Riemannian-manifold-projected Hessian flow—to efficiently compute a solution. Finally, we demonstrate the effectiveness of our approach through examples in urban traffic management, including routing for diverse vehicle types and strategies for minimizing emissions in congested environments.
\end{abstract}

\IEEEpeerreviewmaketitle

\section{Introduction}
In traffic management, each driver—whether operating a car, SUV, or truck—selects the route they perceive to be the shortest. The resulting traffic distribution, in which no driver can unilaterally switch routes to reduce travel time or cost, is known as an equilibrium. For a homogeneous (single-population) flow, this equilibrium is formalized as the Wardrop equilibrium, as stated in the first Wardrop principle \cite{Wardrop_Equilibria}:

\textit{"The journey times on all the routes actually used are equal and less than those which would be experienced by a single vehicle on any unused route."}

While the classical Wardrop model assumes all users have an identical impact on congestion and perceive identical costs, real-world traffic systems are inherently heterogeneous: different vehicle types contribute unequally to congestion and experience different costs. For instance, trucks may slow down traffic more than cars, and their operators may evaluate travel costs differently. To account for these disparities, the Wardrop principle has been extended to multi-population settings, where each population represents a class of users with distinct characteristics and cost functions (see \cite{Uniqueness_general}, \cite{Con_Multi}, and references therein).

Motivated by the need to model more realistic traffic behavior, this paper investigates the multi-population Wardrop equilibrium using a variational inequality framework. This formulation arises naturally from the observation that, at equilibrium, users travel along the shortest paths with respect to the given cost. We discuss the existence and uniqueness of solutions to multi-population Wardrop equilibrium. Furthermore, we introduce a novel method to compute the solution efficiently.

In the literature, the definitions of Wardrop equilibrium and Nash equilibrium are often used interchangeably in the context of non-cooperative network games involving multiple populations. For example, in \cite{Wardrop_Nash}, the authors demonstrate that, under certain conditions defined by entrance-exit pairs and demand patterns, the asymptotic behavior of the Nash–Cournot equilibrium converges to a Wardrop equilibrium. Existence and uniqueness of equilibria are central to network optimization problems. In \cite{Topological_Nash}, it is shown that, in nearly parallel networks, the Nash equilibrium is topologically unique. Conversely, \cite{atomic_not_unique} points out that atomic game equilibria, while existent, are not necessarily unique. Further, \cite{Uniqueness_general} establishes conditions—specifically, diagonal strict convexity (DSC), as derived from \cite{n_games}—that guarantee the existence and uniqueness of a Nash equilibrium for multi-player flows on generalized networks. Inspired by \cite{Uniqueness_general}, the authors in \cite{Polynomial_cost} also derive similar conditions for competitive polynomial cost functions. In this paper, we will employ conditions analogous to those in \cite{Uniqueness_general} and \cite{Survey} to demonstrate the uniqueness of our resulting equilibrium. These DSC conditions can be interpreted as a form of strict monotonicity of the underlying variational inequalities, as will be discussed in Section \ref{Solver}. 

Once the existence and uniqueness of an equilibrium have been established, the next step is to develop an algorithm capable of efficiently computing the equilibrium in multi-player network games. Several computational approaches have been proposed when the equilibrium is characterized via a minimization principle. For example, a Gauss-Seidel approach, which iteratively fixes the flows of all but one player and then optimizes the remaining player's flow using the fixed values from the others \cite{Beckman}. In \cite{TRUCKS_AND_CARS}, the authors show that such an algorithm converges to the equilibrium provided that inter-class interactions are relatively weak compared to the primary effects, evidenced by cost function Jacobian norms being less than one. However, many multi-population problems involve highly coupled cost functions, limiting the applicability of this approach. Furthermore, in \cite{Lemke-like}, the multi-class networking problem is tackled by assuming affine cost functions and reformulating the system as a linear complementarity problem, which is then solved via a Lemke-like algorithm. Despite its theoretical appeal, this method has been found to be unscalable—the computational time increases exponentially with the size of the network and the number of populations. To address this, \cite{Lemke-line-2} extends the Lemke-like algorithm by leveraging properties of hyperplane arrangements, resulting in a polynomial-time solution. However, these methods either cannot be directly applied to our variational case, converge under strong assumptions, or have a high computational and time cost. As such, the multi-population Wardrop equilibrium solver has still been rendered an open problem. In this paper, by leveraging the Hessian Riemannian flow (HRF) (see \cite{Hessian} and \cite{Hessian_MONOTONE}), we develop an algorithm to find multi-population Wardrop equilibrium. Unlike the methods mentioned above, HRF is computationally efficient, globally convergent under mild assumptions, and, most importantly, naturally respects all constraints.

The main contributions of this paper are as follows:
\begin{itemize}
     \item We introduce a new framework for defining the existence and uniqueness of equilibria in multi-population settings.
    \item We propose a novel globally convergent and highly efficient method for solving the multi-population Wardrop equilibrium problem using a Riemannian-manifold-projected Hessian steepest descent approach.
\end{itemize}
The remainder of the paper is organized as follows. In Section \ref{sec:WE}, we present the formulation, existence, and uniqueness results for the Wardrop equilibrium in a single-population setting. Section \ref{sec:Multi_theory} extends these results to the multi-population context. In Section \ref{Solver}, we detail the transformation of the problem into a distributed optimization problem and introduce our Hessian Riemannian flow method. Section \ref{Sec:sims} illustrates simulation results for both non-unique and unique equilibrium scenarios in congestion management, as well as a scenario addressing the minimization of toxic emissions through traffic management. Finally, Section \ref{sec:Conc_Future_work} concludes the paper.

\section{Single-Population Wardrop Model }\label{sec:WE}

The main consideration upon which the model is based on is a steady-state model for agent flow within the network. The model of the Wardrop Equilibrium is formulated as follows:
\begin{itemize}
    \item A  directed graph is given $G=(E,V)$, where  $E=\left \{ e_k,\ :\, k \in \left \{ 1,2,...,n \right\}\right\}$ is the set of edges, while $V=\left \{ v_i,\ :\, i \in \left \{ 1,2,...,m \right\}\right\}$ is the set of vertices. Each edge $e_k$ is associated with a pair of nodes and can be described as $(v_r,v_i)$ where $v_r$ is the standpoint and $v_i$ is the endpoint.
    \item The current $\jmath_k\geq 0$ represents the steady-state flow of agents through the directed edge $e_k$. Let $\jmath=[\jmath_1,\jmath_2, \ldots, \jmath_n]^T$ be the vector that represents all flows across all edges of set $E$.
    \item There exists an associated flow-dependent cost for each edge. Let $\bm{c}=\left \{ c_k,\ :\, k \in \left \{ 1,2,...,n \right\}\right\}$ be the set of cost functions associated with the edges.  Per unit of time corresponding to the distribution of currents $\jmath$ the cost on $e_k$ edge is $ c_{k}(\jmath) \jmath_{k}$ and the total cost is  $\langle c(\jmath), \jmath \rangle:=\sum_{k=1}^{n} c_{k}(\jmath) \jmath_{k}$.
    \item Agents enter through $\lambda$ entrance vertices and exit through $\mu$ exit vertices. It is important to note that the entrance and exit vertices are disjoint (no direct edge joins them).
    \item The flow of agents into the system is described with the entry current, whereas no additional agents are added into the network from other vertices.
\end{itemize}

To describe the gathering and splitting equations of the graph, a Kirchoff matrix $K$ is used, which is an $(m-\mu)\times n$ matrix
defined by
\begin{equation}\label{eq:Kitchoff_build}
K_{i,k} =
\begin{cases}
    1 & \text{if } e_k=(v_i,v_r) \\
    -1 & \text{if } e_k=(v_r,v_i) \\
    0 & \text{if } v_i \notin e_k
\end{cases}
\end{equation}
where $i \in\{1,2, \ldots, m-\mu\}$ and $k \in\{1,2, \ldots, n\}$. The rows of \(K\) correspond to the non-exit vertices of the network, and the columns correspond to the edges. To describe the flow of currents, we construct an \((m - \mu)\)-dimensional vector \(B\), where each component corresponds to a non-exit vertex and represents the fixed, finite inflow into the network.  Thus, the distribution of currents $\jmath$ is admissible, if
\begin{equation} \label{eq:Kirchoff}
    K\jmath=B.
\end{equation}
This is called  Kirchoff law. The set of all admissible distributions of currents is denoted by $\mathcal{A}$:
\begin{equation}
    \mathcal{A}:=\{ \jmath\in\Rr ^n : \jmath \geq \bm{0}, K \jmath=B \}.
\end{equation}

\begin{definition}\label{Def:Wardrop_Eq}
A distribution of currents $\jmath \in \mathcal{A}$ is a Wardrop equilibrium, if $\forall \jmath \in \mathcal{A}$:
\begin{equation}\label{eq:Optimal_variation}
\left\langle c(\jmath^{*}), \jmath^{*}-\jmath \right\rangle\leq 0 .
\end{equation}
\end{definition}
For the existence of solutions, see, for example, Theorem 3.1 in \cite{Var_prob}.

\begin{theorem}\label{theorem-exist-single}
    Let $c$ be a continuous mapping from $\mathcal{A}$ into $\mathbb{R}^n$. Then, there exists a Wardrop Equilibrium.
\end{theorem}
\begin{proof}
   Noticing that $\mathcal{A}$ is nonempty, compact (due to constant vector $B$), and convex subset of $\mathbb{R}^n$ by Theorem 3.1 in \cite{Var_prob}, we conclude the proof. 
\end{proof}

The notion of monotonicity is used to prove the uniqueness of Wardrop Equilibrium. 
\begin{definition}
    A cost $c(\jmath)$ is monotone, if $\forall \jmath_1,\jmath_2 \in \mathcal{A}$,
    \begin{equation*}\label{eq:Wardrop_monotone}
        \left\langle c(\jmath_1)-c(\jmath_2), \jmath_1-\jmath_2 \right\rangle\geq 0 .
    \end{equation*}
    If $\jmath_1 \neq \jmath_2$, the inequality is strict; the cost $c(\jmath)$ is called strictly monotone.
\end{definition}
This strict monotonicity is a sufficient condition for uniqueness. See \cite{Var_prob} or \cite{Wardrop1}. 
\begin{theorem}
Assume that the cost $c(\jmath)$ is strictly monotone on $\mathcal{A}$. Then, there is at most one Wardrop Equilibrium. 
\end{theorem}

\section{Multi-Population Wardrop Model}\label{sec:Multi_theory}

In this section, we define and analyze the multi-population Wardrop model. 

Consider a network represented by a directed graph $G$ (see Section \ref{sec:WE}) with multiple entrances and exits.   The multi-population model is described as follows:

\begin{itemize} \item The total population of agents $\Lambda$ is composed of $P$ distinct sub-populations, denoted by $\Lambda_1, \dots, \Lambda_P$.
    \item For each population $\Lambda_r$, where $r \in \{1,\dots,P\}$, we denote its current (flow) distribution by $\jmath^r = (\jmath^r_1, \dots, \jmath^r_n) \in \mathbb{R}^n_{\geq 0}$, where $\jmath^r_k$ represents the flow of population $r$ on edge $e_k$. For a fixed edge $e_k$, the population-wise flow vector is $J_k = (\jmath^1_k, \dots, \jmath^P_k)$.

\item Each population $\Lambda_r$ enters through a set of $\lambda_r$ entrance vertices and exits through a set of $\mu$ exit vertices. The total number of unique entrance vertices across all populations is at most $\lambda = \sum_{r=1}^P \lambda_r$, noting that some entrance vertices may be shared among populations. As in the single-population case, entrance and exit vertices are assumed to be disjoint (no direct edge connects them).

\item No agents are introduced or removed from the network at vertices other than the designated entrance and exit vertices. This constraint yields the splitting and gathering equations (Kirchhoff’s law) for each population:
\[
K \jmath^r = B^r, \quad r = 1,\dots,P,
\]
where $K$ is the Kirchhoff matrix (see Section~\ref{sec:WE}) and $B^r \in \mathbb{R}^{m-\mu}$ is the net input vector for population $r$.
 The compact form of the flow balance condition for all populations is:
\begin{equation}\label{def-multi-Kr}
    K J = \bm{B},
\end{equation}
where $J := (\jmath^1 \,|\, \cdots \,|\, \jmath^P) \in \mathbb{R}^{n \times P}$ and $\bm{B} := (B^1 \,|\, \cdots \,|\, B^P) \in \mathbb{R}^{(m-\mu) \times P}$.

\item The admissible set of current distributions for population $\Lambda_r$ is defined as:
\[
\mathcal{A}^r := \{ \jmath^r \in \mathbb{R}^n_{\geq 0} : K \jmath^r = B^r \},
\]
and the admissible set for the entire system is the Cartesian product:
\[
\bm{\mathcal{A}} := \mathcal{A}^1 \times \cdots \times \mathcal{A}^P.
\]
Hence, $J$ is admissible, i.e. $J\in \bm{\mathcal{A}}$,   if it 
 satisfies \eqref{def-multi-Kr}. 

\item Each population may have a distinct contribution to the cost on each edge, which may depend on the flows of all populations.  For each population $r$, its cost vector is $c^r(J) = (c^r_1(J), \dots, c^r_n(J))$, where each component $c^r_k(J)$ reflects the cost per unit flow for $\Lambda_r$ on edge $e_k$, influenced by the full vector of edge flows $J_k$.

\item The combined cost on edge $e_k$ across all populations is denoted $c_k(J) = (c^1_k(J), \dots, c^P_k(J)) $, and the system-wide 
cost profile can be compiled as $C(J) = (c_1(J), \dots, c_n(J))$.

\end{itemize}

Now, based on the discussion above, we define the \textit{social cost} (i.e., total cost) of the entire population $\Lambda$, which leads to the notion of a \textit{Wardrop equilibrium}. In this case, all populations collectively aim to minimize the overall cost in the network, and the equilibrium corresponds to a solution of a global optimization problem.

So, the total cost per unit of time is 
\begin{equation}\label{def-cost-social}
    \langle C(J), J \rangle: =\sum_{r=1}^{P} \langle c^r(J), \jmath^r\rangle=\sum_{r=1}^{P} \left( \sum_{k=1}^{n}  
 \jmath^r_k c^r_k (J) \right).
\end{equation}

\begin{definition}\label{Def:Wardrop}[Wardrop Equilibrium]
We  say that $\bar{J} \in \bm{\mathcal{A}}$ is a Wardrop Equilibrium of the  multi-population system if for all $ J \in \bm{\mathcal{A}}$,
\begin{equation}\label{eq:OPT_VAR_CEN_MULTI}
\left\langle C(\bar{J} ), \bar{J}-J \right\rangle\leq 0 .
\end{equation}
\end{definition}
Similarly to Theorem \ref{theorem-exist-single}, we have the existence of Wardrop equilibrium.
\begin{theorem}\label{thm:wardrop_existence}
    Let $C(\cdot)$ be a continuous mapping from $\bm{\mathcal{A}}$ into $\mathbb{R}^{n\times P}$. Then, there exists a Wardrop Equilibrium.
\end{theorem}

The notion of monotonicity is used to prove the uniqueness of Wardrop equilibrium. 
\begin{definition}\label{def:Monotonicity}
    A cost of a the general multi-population system $C(J)$ is monotone if for all values $J_1 ,J_2 $ that belong to  $\bm{\mathcal{A}}$,
    \begin{equation}\label{eq:MONOTONE_MULTI_CENT}
        \left\langle C(J_1)-C(J_2), J_1-J_2 \right\rangle\geq 0 .
    \end{equation}
    If $J_1 \neq J_2$, the inequality is strict; in this case, we can say that the cost $C(J)$ is strictly monotone.
\end{definition}
If each population's cost is monotone, then the total cost is also monotone.

This strict monotonicity is a sufficient condition for uniqueness. 
\begin{theorem}\label{thm:Wardrop_uniqueness}
Assume that the cost $C(\cdot)$ is strictly monotone on $\bm{\mathcal{A}}$. Then, there is at most one  Wardrop Equilibrium. 
\end{theorem}
\begin{proof}
    Assume $J_1$ and $J_2$ are both Wardrop Equilibria. This means that for all $J \in \bm{\mathcal{A}}$ we have:
\begin{equation}
    \left\langle C(J_1), J_1-J \right\rangle\leq 0, \; \text{and} \; \left\langle C(J_2), J_2-J \right\rangle\leq 0.
\end{equation}
Accordingly,
\begin{align*}
    \left\langle C(J_1)-\Tilde{{C}}(J_2), J_1-J_2 \right\rangle \leq 0,
\end{align*}
and since $C$ is strictly monotone, we deduce that $J_1=J_2$. 
\end{proof}

In the next section, we present a numerical method to find the multi-population Wardrop equilibrium.

\vspace{-2mm}
\section{Multi-Population Wardrop Solver}\label{Solver}
To find the solution to the multi-population problem, first, we reformulate the problem as a coordinate-wise problem or population-wise problem. Then, we show that under the monotonicity condition, the Wardrop equilibrium coincides with the solution of the population-wise problem. Finally, we develop a numerical method for the population-wise problem.

\subsection{Problem Reformulation}

The population-wise cost per unit of time is 
\begin{equation}\label{def-cost-social}
\langle c^r(J), \jmath^r \rangle: =   \sum_{k=1}^{n}  
 \jmath^r_k c^r_k (J).
\end{equation}
Note that the proceeding coincides with the single population Wardrop model; the only difference is that the cost functions not only depend on $\jmath^r$ but also on $\jmath^i$, $i=1,\dots,(r-1),(r+1),\dots, \jmath^P$. 
Now, using this notation, we define the population-wise problem, which is the Nash equilibrium between the populations with the common cost function.

\begin{definition}
A collection of flows $J^*=\{\jmath^{r*}\}_{r=1}^P \in \bm{\mathcal{A}}$ is a \textit{Nash equilibrium} or solution to population-wise problem if, for all populations $r \in \{1,\dots,P\}$, the following variational inequality holds:
\begin{equation}\label{def-NE}
    \left\langle c^r(J^*), \jmath^{r*} - \jmath^r \right\rangle \leq 0,
\end{equation}
for all $ \jmath^r \in \mathcal{A}^r$.
\end{definition}

\begin{theorem}\label{theorem-NCW}
The population-wise (Nash) and Wardrop problems are equivalent. 
\end{theorem}
\begin{proof}We prove that  Wardrop equilibrium is a population-wise solution, too, and the opposite of each population-wise solution is also a Wardrop equilibrium.

  Let $\bar{J}\in\bm{\Aa}$ is a Wardrop equilibrium; that is, \eqref{eq:OPT_VAR_CEN_MULTI} holds for   $\bar{J}\in\bm{\Aa}$ and any $J\in\bm{\Aa}$. Now, in  \eqref{eq:OPT_VAR_CEN_MULTI} taking $J^i=\bar{J}^i$ for all $\in\{1,2,\dots,P\}$ and $i\neq r$ by \eqref{def-cost-social}, we 
  get
  \begin{equation*}
    \left\langle c^r(\bar{J}), \bar{\jmath}^{r} - \jmath^r \right\rangle \leq 0.
\end{equation*}
  Hence, $\bar{J}$ is a population-wise solution. 

  Suppose $J^*\in\Aa$ is a population-wise solution. Then, adding up \eqref{def-NE} for all $r=1,\dots,P$, we obtain \eqref{eq:OPT_VAR_CEN_MULTI}.
\end{proof}

\subsection{Approach}
Using population-wise formulation,  we propose a numerical solver based on Hessian-Riemannian flow (HRF), \cite{Hessian}. Unlike classical Euclidean methods, HRF naturally respects boundaries and constraints through the metric itself—no need for external projection steps. Furthermore, the method is globally convergent, and we adapted to find Wardrop equilibrium.

In \cite{Hessian}, the authors propose a solution for solving a minimization problem under equality constraint in a convex and open set. The proposed problem solution consists of endowing the open convex set with a Riemannian structure. This restricts the search to the relative interior of the feasible set defined by the linear equality. By applying the Hessian steepest descent method, orthogonally projected onto the Riemannian manifold, one obtains a projected gradient descent solution for the constrained problem. In this formulation, the steepest descent method becomes a local minimization process on a Riemannian manifold, where a vector field generates solution trajectories based on initial conditions from the interior of the domain. The approach in \cite{Hessian} is called the Hessian-Riemannian gradient flow (HRGF), which is constructed using the cost function gradient. 
In \cite{Hessian_MONOTONE}, in the context of mean-field games, the authors demonstrated that the gradient of the cost function in the HRGF approach can be replaced with monotone operators.
Motivated by this, we use Hessian-Riemannian flow to solve the system of variational inequalities 
 defined by the multi-population Wardrop Equilibrium.

We start by considering the single population problem in \eqref{def-NE} with fixed 
$\jmath_F^{-r} := (\jmath_F^1 , \dots, \jmath_F^{r-1},\jmath_F^{r+1}, \dots, \jmath_F^{P}) \in \mathbb{R}^{n \times {(P-1)}}$ and aim to find $\bar{\jmath}^{r}\in \Aa^r$ such that for all $\jmath^{r}\in \Aa^r$ the following inequality holds
\begin{equation}\label{def-single-Wardrop}
    \left\langle c^r(\bar{\jmath}^{r},\jmath_F^{-r}), \bar{\jmath}^{r}- \jmath^{r}) \right\rangle \leq 0.
\end{equation}
\vspace{-9mm}
\begin{remark}
    In some cases, the population $r$ may occupy some part of the original graph $G$. Particularly, if there are total $\lambda_{T}$ entry vertices, the $r$ population agents enter the network only when $\lambda_r<\lambda_{T}$. In this case, the $r$-th population's    current  on the remaining $(\lambda_{T}-\lambda_r)$ entry vertices is $0$. 
\end{remark}
\vspace{-2mm}
To initiate the HRF method, we require an interior point, \( j^r > 0 \). Hence, if the \( r \)-th population always has zero current on some edge, we cannot properly initialize the HRF. Therefore, if it is known \emph{a priori} that the \( r \)-th population always has zero current on certain \( \lambda_{-r} \) edges, we remove those edges from the original graph. This results in a subgraph \( G_r \), along with the corresponding Kirchhoff matrix \( K_r \) and \( B_R^r \), which corresponds to a non-exit vertex. This process does not affect the solution and ensures that we can select an interior point required to initialize the HRF. 

Now, relying on the discussion above, we reformulate the problem in \eqref{def-single-Wardrop}.
So, instead of original distribution of currents $\jmath^r\in R^n$, we consider its sub-vector    $\vartheta^r\in R^{n-\lambda_{-r}}$ (all constant $0$ components are removed). The admissible set for each population becomes $\Aa_R^r:=\{ \vartheta^r\in \jmath^r\in R^{n-\lambda_{-r}}: \vartheta^r\geq 0,\,\ K_r \vartheta^r =B_R^r \}$. Thus,
the population-wise problem reads as follows:
\vspace{-4mm}
\begin{problem}\label{prob-ax} For fixed 
$\vartheta^{-r} := (\vartheta^1 , \dots, \vartheta^{r-1},\vartheta^{r+1}, \dots, \vartheta^{P}) \in \mathbb{R}^{n \times {(P-1)}-\sum_{k\neq r}\lambda_{-k}}$ find $\bar{\vartheta}^{r}\in \Aa_R^r$ such that for all $\vartheta^{r}\in \Aa_R^r$, the following inequality holds
\begin{equation}\label{def-sub-single-Wardrop-}
    \left\langle c^r(\bar{\vartheta}^{r},\vartheta^{-r}), \bar{\vartheta}^{r}- \vartheta^{r}) \right\rangle \leq 0.
\end{equation}
\end{problem}

For fixed 
$\vartheta^{-r} = (\vartheta^1 , \dots, \vartheta^{r-1},\vartheta^{r+1}, \dots, \vartheta^{P})$, we find a solution to Problem \ref{prob-ax} by HRF method, which leads to the solution of an ODE. Therefore, to find a population-wise problem solution for all $\vartheta^{-r} = (\vartheta^1, \dots, \vartheta^{r-1},\vartheta^{r+1}, \dots, \vartheta^{P})$, $r=1,\dots, P$, we write the corresponding ODE and end up with a coupled system of ODEs, the solution of which -- due HRF method -- converges to the solution of our population-wise problem, later shown to be Wardrop equilibrium.  

For our main result, we consider the following:
\[
h_r(x) = x \log x = \sum_{k=1}^{n - \lambda_{-r}} x_k \log x_k,
\]
Legendre-type strictly convex function on \( \mathbb{R}^{n - \lambda_{-r}}_{\geq 0} \) (see Definition 3.1 in \cite{Hessian}).
Let \( H_r(x) = \nabla^2 h_r(x) \) and set
\begin{equation}\label{eq-flow}
F_r(\vartheta^r)=\left[I-{K}_r^{T}G_r(\vartheta^r)\right],
\end{equation}
 where 
 \begin{equation*}
   G_r(\vartheta^r)=  \left(K_r H_r(\vartheta^r)^{-1} {K}_r^{T}\right)^{-1} K_r H_r(\vartheta^r)^{-1}.
 \end{equation*}
 For the proof, we use the \( h \)-divergence (Bregman divergence) function associated with \( h_r \), defined as
\begin{equation*}
    d_{h_r}(x, y) = h_r(x) - h_r(y) - (x - y) \cdot \nabla h_r(y),
\end{equation*}
which is non-negative and  strictly positive whenever \( x \neq y \).
 
\begin{theorem}\label{theorem-main}
    Let the cost functions $c^r(\cdot)$ be continuous, locally Lipchitz and strictly monotone. Suppose that  $\vartheta(0)=\vartheta_0>0$, such that $\theta^r_0 \in \mathcal{A}^r_R$, and $\vartheta (t)=(\vartheta^1(t),\dots,\vartheta^P(t))$ solves the following system of ODEs 
\begin{equation}\label{theorm-eq-1}
\begin{cases}
\dot{\vartheta}^r+H_r(\vartheta^r)^{-1}F_r(\vartheta^r)c^r(\vartheta^r,\vartheta^{-r}) =0 \\
\vartheta^r(0)=\vartheta^r_0, 
\end{cases}\,\ r=1,\dots, P,
    \end{equation}
  where $F_r$ is defined by \eqref{eq-flow}.
    Then, $\vartheta^r(t) \in \mathcal{A}^r_R$ for all $ t \geq 0$. Also, there exists $\lim_{t\to\infty}\vartheta(t)=\vartheta_\infty$, and $\jmath_\infty$, corresponding to $\jmath_{\infty}$, is the Wardrop equilibrium. 
\end{theorem}
\begin{proof}
Note that it is enough to prove that there exists  $\lim_{t\to\infty}\vartheta(t)=\vartheta_\infty$, and $\vartheta_\infty$ solves Problem \ref{prob-ax}.

We first prove the well-posedness of \eqref{theorm-eq-1}. The theorem assumptions with the definition of  \( H_r \) imply that the right-hand side of \eqref{theorm-eq-1} is locally Lipschitz in \( \vartheta \). Therefore, by the Picard--Lindel{\"o}f theorem, solutions' local existence and uniqueness follow. 

Next, we prove the global existence and uniqueness of the solutions. 
Let $\{\vartheta^r(t)\}_{r=1}^{P}$ be a solution to \eqref{theorm-eq-1}.  Note that
    \begin{equation*}
    \begin{split}   &K_rH_r(\vartheta^r)^{-1}F_r(\vartheta^r)=K_rH_r(\vartheta^r)^{-1}\\&-K_rH_r(\vartheta^r)^{-1}K_r^T(K_rH_r(\vartheta^r)^{-1}K_r^T)^{-1}K_rH_r(\vartheta^r)^{-1}=0,
    \end{split}
    \end{equation*}
    which, along with \eqref{theorm-eq-1}, implies that $K_r\dot{\vartheta}^r(t)=0$. Using this and  
    recalling that $K_r \vartheta(0)=K_r\vartheta_0 =B_r$, we get
   \begin{equation} \label{theorem-proof-eq3}
K_r\vartheta^r(t)=\int_{0}^{t} K_r\dot{\vartheta}^r(s) d s+K_r \vartheta(0)=K_r\vartheta_0 =B_r.
   \end{equation}
   
By Theorems \ref{thm:wardrop_existence}, \ref{thm:Wardrop_uniqueness}, \ref{theorem-NCW}, we deduce that there exists a unique solution to Problem \ref{prob-ax}. Let $\bar{\vartheta}$, $\bar{\vartheta}^r\in \mathcal{A}^r_R$ be the unique solution to Problem \ref{prob-ax}.
Hence, by \eqref{theorem-proof-eq3}, we have
\begin{equation}\label{theorem-proof-eq4}
    \left(\bar{\vartheta}^r-\vartheta^r(t)\right)^T{K}_r^{T}=0.
\end{equation}

Because \( c^r \) is strictly monotone and \( \bar{\vartheta} \) is the solution to Problem~\ref{prob-ax}, by combining \eqref{eq:MONOTONE_MULTI_CENT} and \eqref{def-sub-single-Wardrop-}, we obtain 
\begin{equation*}
\begin{split}
    \left\langle c^r(\vartheta), \vartheta^r-\bar{\vartheta}^r\right\rangle &=\left\langle c^r\left(\bar{\vartheta}^{r}, \vartheta^{-r}\right)-c^r\left(\vartheta^{r}, \vartheta^{-r}\right), \bar{\vartheta}^{r}-\vartheta^r\right\rangle\\&-\left\langle c^r\left(\bar{\vartheta}^{r}, \vartheta^{-r}\right), \bar{\vartheta}^{r}-\vartheta^r\right\rangle > 0.
\end{split}
\end{equation*}
Taking $\vartheta=\vartheta(t)$ in the proceeding equation, we get
\begin{equation}
    \label{theorem-proof-eq1}
    \left\langle c^r(\vartheta(t)), \bar{\vartheta}^r-\vartheta^r(t)\right\rangle =\left(\bar{\vartheta}^r-\vartheta^r(t)\right)^Tc^r(\vartheta(t))< 0.
\end{equation}
Recalling that $h^r$ is strictly convex function for $\bar{\vartheta}^r\neq\vartheta^r(t)$, we have
\begin{equation}\label{theorem-proof-eq2}
\begin{split}
    d_{h_r}&(\bar{\vartheta}^r,\vartheta^r(t))=h_r(\bar{\vartheta}^r)-h_r(\vartheta^r(t))\\&-(\bar{\vartheta}^r-\vartheta^r(t))^{T} \nabla h_r(\vartheta^r(t))\\&=h_r(\bar{\vartheta}^r)-(\bar{\vartheta}^r)^{T}\log\vartheta^r(t)+\vartheta^r(t)- \bar{\vartheta}^r> 0.
\end{split}
\end{equation}
Next, we show that $d_{h_r}(\bar{\vartheta}^r,\vartheta^r(t))$ strictly decreases.
To do so, we consider the time derivative of $d_{h_r}(\bar{\vartheta^r},\vartheta^r(t))$
\begin{equation*}
   \begin{split}
        \frac{d }{dt}d_{h_r}(\bar{\vartheta^r},\vartheta^r(t))&=-\dot{\vartheta^r}^{T}(t) \nabla h_r(\vartheta^r(t))+\dot{\vartheta^r}^{T}(t) \nabla h_r(\vartheta^r(t))\\&-\left(\bar{\vartheta}^r-\vartheta^r(t)\right)^{T} H_r(\vartheta^r(t)) \dot{\vartheta^r}(t) 
 \\&=-\left(\bar{\vartheta}^r-\vartheta^r(t)\right)^{T} H_r(\vartheta^r(t)) \dot{\vartheta^r}(t). 
   \end{split}
\end{equation*}
Using expression of $\dot{\vartheta^r}(t)$ from \eqref{theorm-eq-1} in the proceeding equation, by equations in \eqref{theorem-proof-eq4}   and \eqref{eq-flow}, we get
\begin{equation}\label{theorem-proof-eq5}
   \begin{split}
        \frac{d }{dt}&d_{h_r}(\bar{\vartheta^r},\vartheta^r(t))=\left(\bar{\vartheta}^r-\vartheta^r(t)\right)^Tc^r(\vartheta(t))\\&+\left(\bar{\vartheta}^r-\vartheta^r(t)\right)^T{K}_r^{T} G_r(\vartheta^r(t))c^r(\vartheta(t))\\&=\left(\bar{\vartheta}^r-\vartheta^r(t)\right)^Tc^r(\vartheta(t))<0,
   \end{split}
\end{equation}
where the last inequality follows from \eqref{theorem-proof-eq1}. 
On the other hand, note that for any $a>0$ there exists $M>0$ such that
\begin{equation*}
    |\{ d_{h_r}(x,y)\leq a: x\in \mathbb{R}_{\geq 0}^{n - \lambda_{-r}},\,\ y\in \mathbb{R}_{ > 0}^{n - \lambda_{-r}}\}| \leq M.
\end{equation*}
Along with
the equations \eqref{theorem-proof-eq2} and \eqref{theorem-proof-eq5}, it  implies that $\vartheta^r(t)$ is bounded. Therefore, by Theorem 3.3 in \cite{khalil2002nonlinear}, we deduce that  for all $t\geq0$ there exists a unique solution to  \eqref{theorm-eq-1}, defined as $\{\vartheta^r(t)\}_{r=1}^{P}$. Furthermore, there exists
$\vartheta_\infty\geq 0$ such that $\lim_{t\to\infty}\vartheta(t)=\vartheta_\infty$. Next, we prove that \(\vartheta_\infty\) solves Problem~\ref{prob-ax}. Note that Equations~\eqref{theorem-proof-eq2} and \eqref{theorem-proof-eq5}, together with the non-negativity of the function \(d_{h_r}\), imply that \(\bar{\vartheta}^r\) is an equilibrium point of Equation~\eqref{theorm-eq-1}. Furthermore, using these equations again, we deduce that the function \(d_{h_r}(\bar{\vartheta}^r, \cdot)\) is a Lyapunov function. Therefore, by \cite[Theorem 4.2]{khalil2002nonlinear}, the solution \(\{\vartheta^r(t)\}_{r=1}^{P}\) is globally convergent to \(\bar{\vartheta}\).
\end{proof}

\vspace{-2mm}
\section{Simulation Results}\label{Sec:sims} 
\vspace{-2mm}
In this section, we present three simulation scenarios.
In the first scenario, we validate our multi-population approach by artificially dividing a single population into two sub-populations. We then compare our solution with well-known methods to ensure consistency and accuracy.
The second scenario explores the effects of multi-population congestion. Specifically, we consider populations with different weights—for example, in a traffic management context where trucks experience twice the congestion cost of cars.
In the third scenario, we demonstrate how the multi-population model can be naturally applied to minimize overall emissions using traffic management techniques.
\vspace{-3mm}

\subsection{Scenario 1} In this part, to validate our approach, we compare a uniform population nonlinear programming (NLP) solver, a uniform single-population method using the Hessian Riemannian flow, and a multi-population approach with equal weights solved via the Hessian Riemannian flow. The comparison demonstrates that all three methods yield the same solution.  However, the HRF is computationally efficient. 

\begin{figure*}[t!]
\vspace{-1mm}
    \begin{subfigure}[t]{0.35\textwidth}
        \centering
\includegraphics[width=\textwidth]{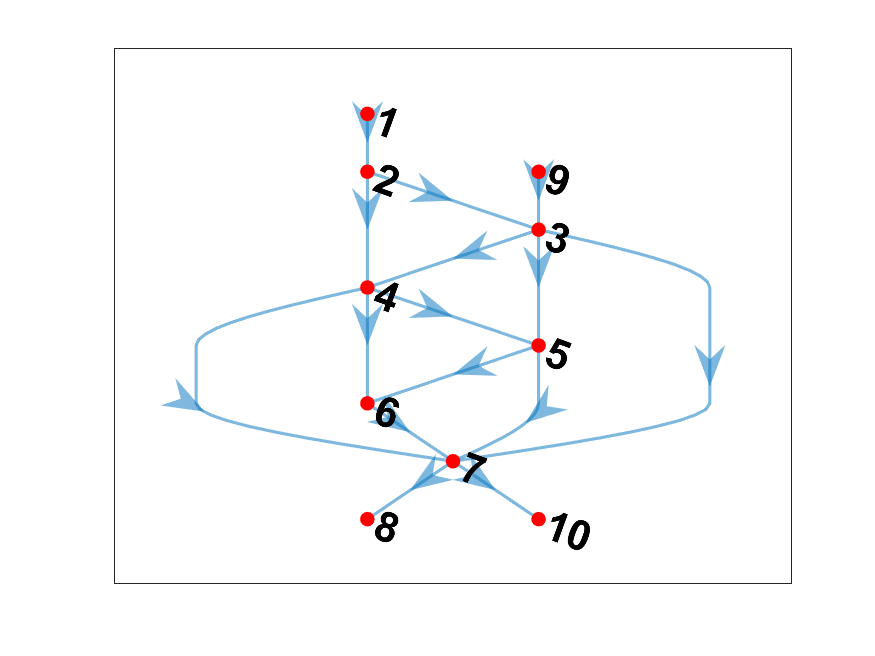}
    \caption{Full road graph representing all roads.}
    \label{fig:FULL_GRAPH}
    \end{subfigure}
    ~
    \begin{subfigure}[t]{0.35\textwidth}
        \centering
        \includegraphics[width=\textwidth]{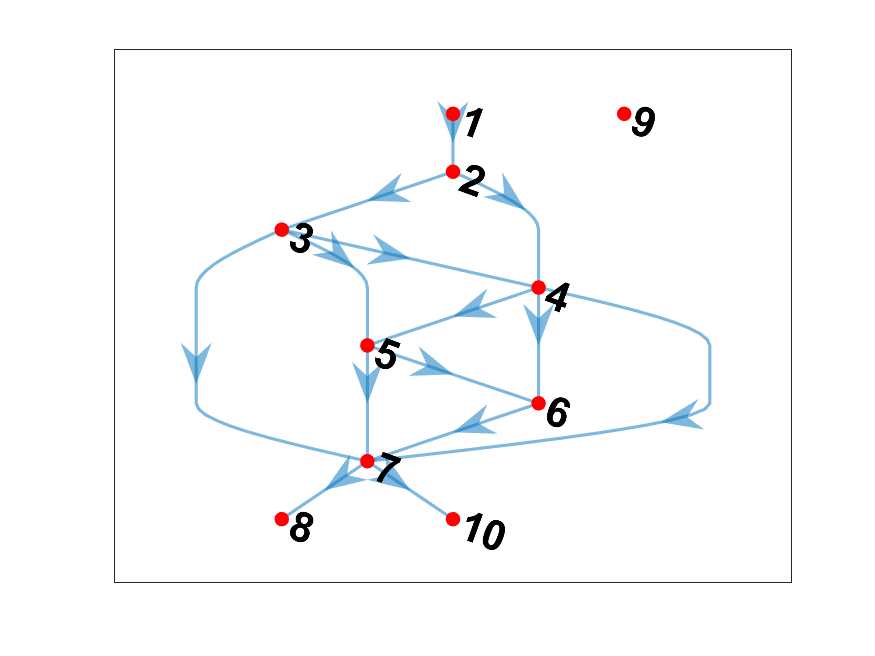}
        \caption{Population $1$ flow sub-graph.}
        \label{fig:Pop1}
    \end{subfigure}%
    ~
    \begin{subfigure}[t]{0.35\textwidth}
        \centering
\includegraphics[width=\textwidth]{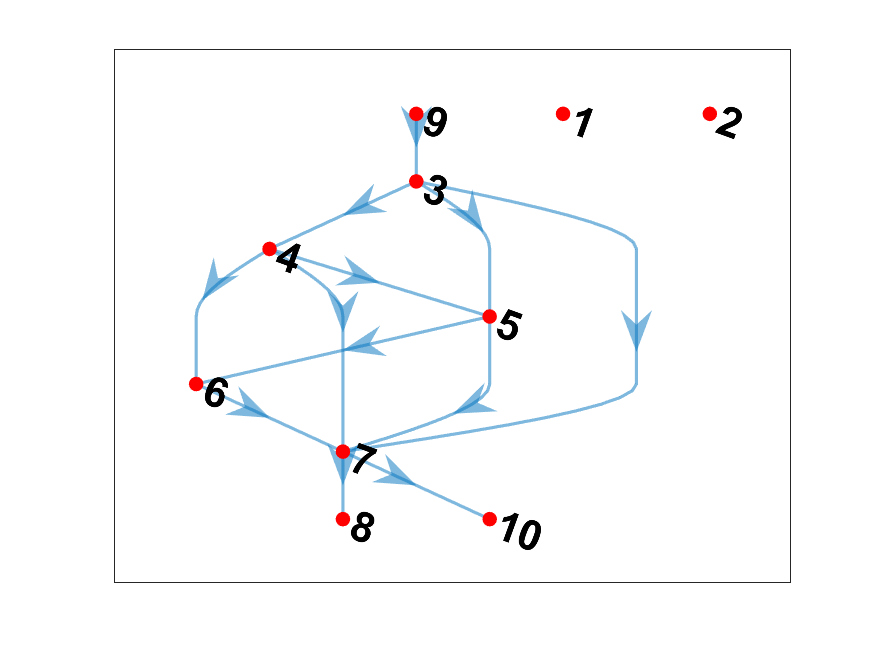}
        \caption{Population $2$ flow sub-graph.}
        \label{fig:Pop2}
    \end{subfigure}    
    \caption{Individual Population sub-graphs.}
    \vspace{-1mm}
\end{figure*}

We consider a case with two populations having equal weights. In this simple scenario, the cost on each edge is defined as the sum of the flows from both populations: $   c_k(\jmath_k) = \jmath_k^1 + \jmath_k^2$,
where \(\jmath_k^i\) denotes the flow of population \(i\) on edge \(k\). The underlying road network is modeled by the connected graph shown in Figure~\ref{fig:FULL_GRAPH}.
Each population utilizes only a portion of the overall network. In particular, the flow \(\jmath^1\) (population $1$) enters exclusively at node $1$, while \(\jmath^2\) (population $2$) enters exclusively at node $9$.

\subsubsection{Uniform Population}
Under the uniform population assumption, we aggregate the flows as \(\jmath = \jmath^1 + \jmath^2\) and define the cost on edge \(k\) as 
\(
c_k = \jmath_k,
\)
where \(\jmath_k\) denotes the total flow on edge \(k\). The edge variables are defined accordingly.
Considering that the inflows at nodes $1$ and $9$ are both $100$, the net input vector for the Kirchhoff equation \(K\jmath = B_0\) is given by
\(
B_0 = [100 , 0 , 0 , 0 , 0 , 0 , 0 , 100]^T.
\)
     Simulating equation~\eqref{theorm-eq-1} for the single-population case achieves convergence in a single time step; the resulting flow values are presented in Table~\ref{tab:DIST_RES}. Note that the flow values are rounded to the nearest integer.
     
     \subsubsection{Nonlinear Programming}
Alternatively, since the cost function is linear—and therefore the gradient of a potential function—the uniform population flow problem can be formulated as the following constrained optimization problem:
\begin{gather*} 
    \min \; \frac{1}{2}\sum_{i=1}^{15}\jmath_i^2,\quad
    \text{s.t.} \; K\jmath = B_0, \;\jmath_i \geq 0,\quad \forall\, i \in \{1, \dots, 15\}.
\end{gather*}
The same solution is obtained using a conventional interior-point optimization toolbox (see Table~\ref{tab:DIST_RES}). However, the conventional solver needed more than $30$ minutes of computational time, rendering it unscalable for larger problems.
\subsubsection{Multi-population Solution}
The uniform population approach does not differentiate between the contributions of individual populations. Therefore, we introduce a multi-population formulation. This formulation defines separate flow variables for each population on each edge. For example, population $1$ flows from entrance node $1$ to nodes $8$ and $10$, while population $2$ flows from entrance node $9$ to nodes $8$ and $10$. Figures~\ref{fig:Pop1} and \ref{fig:Pop2} illustrate the subgraphs corresponding to populations $1$ and $2$, respectively.

 The inflows for both populations are $100$, and for each population  the Kirchhoff law is $K_r \jmath_r=B_r$ with
\[
B_1 = [100,0,0,0,0,0,0]^T, \;
B_2=[0,0,0,0,0,100]^T.
\]
Corresponding cost functions are 
$
c_k(J) = \jmath_k = \jmath_k^1 + \jmath_k^2$.
Because the HRF is globally convergent, we initialize the system of ODE in ~\eqref{theorm-eq-1} with an initial guess for \(J\).  Then, we used 'ode3' in Simulink\textsuperscript{\textregistered} along with a pseudo-inverse block to propagate the new system of ODE's through time. The system converged to the optimal solution in less than $0.2$ seconds.

In this case as well, we obtained the same results as shown in Table~\ref{tab:DIST_RES}. However,  the distributed optimization for the multi-population and the uniform population methods give the solution less than a second.   

\begin{table*}[h]
\centering

    \begin{tabular}{|c|c|c|c|c|c|c|c|c|c|c|c|c|c|c|c|}
    \hline
      Edges   & (1,2) & (2,3) & (9,3) & (2,4) & (3,4) & (3,5) & (4,5) & (4,6) & (5,6) & (3,7) & (4,7) & (5,7) & (6,7) & (7,8) & (7,10) \\
    \hline
    \hline
    Flow 1 & 100   & 38    & 0     & 62    & 1     & 13    & 10    & 15    & 4     & 24    & 39    & 19    & 18    & 50   & 50     \\
    \hline
    Flow 2 & 0     & 0     & 100   & 0     & 23    & 24    & 2     & 7     & 6     & 52    & 13    & 21    & 13    & 50     & 50   \\
\hline
    Total Flow & 100   & 38    & 100   & 62    & 24    & 37    & 12    & 22    & 10    & 76    & 54    & 40    & 31    & 100   & 100   \\
    \hline
    \end{tabular}
    \caption{The table of flow solutions from uniform population, multi-population, and NLP solutions.}
    \label{tab:DIST_RES}
\end{table*}
\vspace{-1mm}
\subsection{Scenario 2}
Next, we consider the same graph as in Figure~\ref{fig:FULL_GRAPH}, with the corresponding subgraphs shown in Figures~\ref{fig:Pop1} and~\ref{fig:Pop2}. In this scenario, population~$1$ represents cars, and population~$2$ represents trucks. We assume that trucks incur twice the congestion cost of cars.
In the case of linear cost, we obtain the same congestion pattern as in Scenario~$1$. This result is expected: compared to Scenario~$1$, the number of vehicles in population~$1$ is reduced by half, but the increased cost weight offsets their reduced contribution to congestion.

\begin{figure}[t!]
\centering
    \begin{subfigure}[t]{0.4\textwidth}
        \centering
        \includegraphics[width=\textwidth]{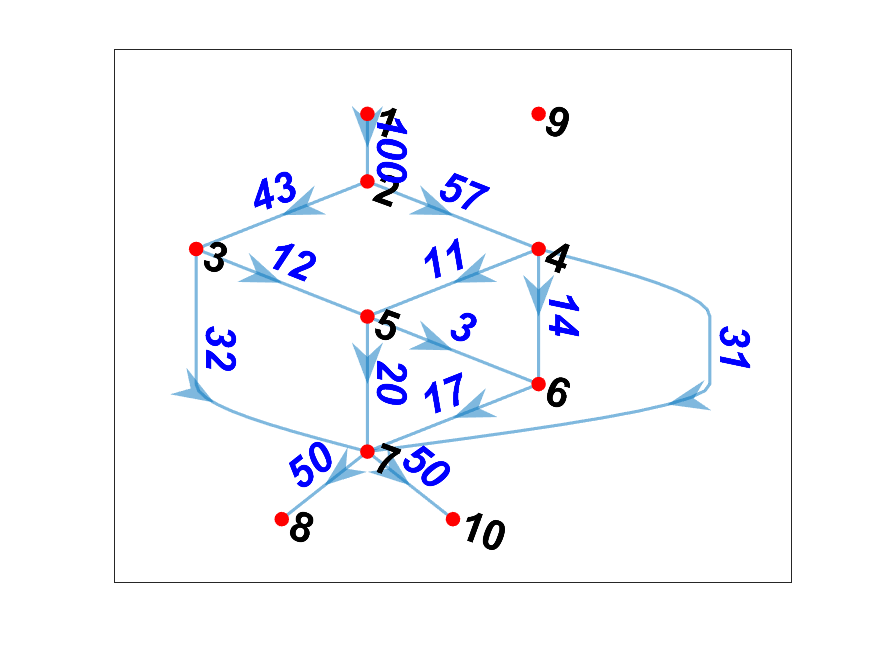}
        \caption{Car flow sub-graph}
        \label{fig:Pop1_scenario2}
    \end{subfigure}%
    \\
    \begin{subfigure}[t]{0.4\textwidth}
        \centering
        \includegraphics[width=\textwidth]{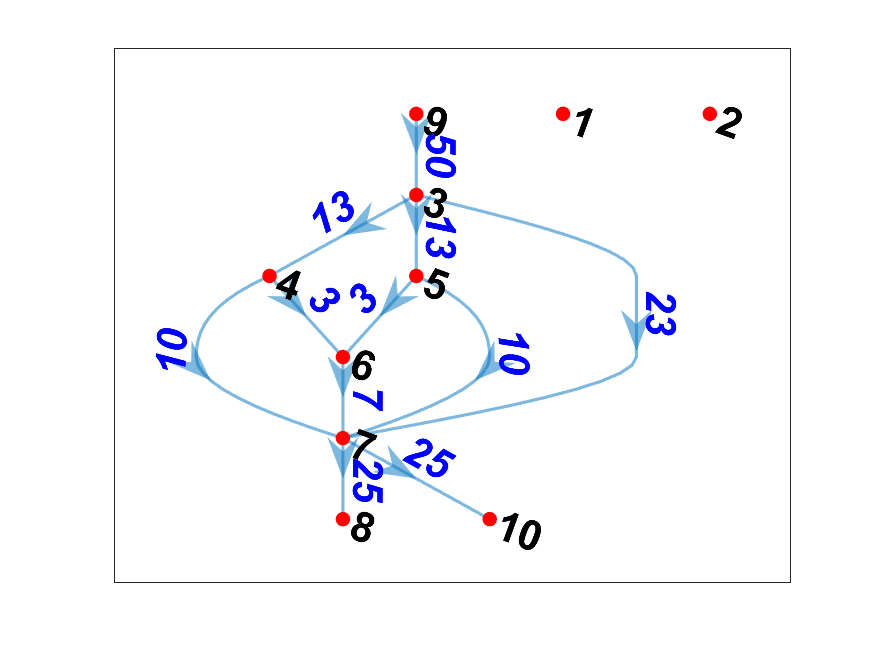}
        \caption{Truck flow sub-graph}
        \label{fig:Pop2_scenario2}
    \end{subfigure}    
    \caption{Individual Population sub-graphs}
    \label{fig:DIST_DIF_COST}
\end{figure}

With the multi-population model, we can impose additional objectives: not only do we aim to minimize congestion via optimal density flow, but we also seek to distribute the presence of trucks and cars more evenly across the grid. These two objectives are encoded in the edge cost function for each population, defined as
$c_k^r(\jmath_k) = 0.5 (\jmath_k^1 + 2\jmath_k^2) + 0.5\jmath_k^r$.
This cost function satisfies the strict monotonicity conditions required by Theorem~\ref{thm:Wardrop_uniqueness}, resulting in a unique equilibrium. Additionally, only $50$ trucks enter from node $9$, while $100$ cars enter from node $1$. 
 The resulting flow distributions, obtained by solving Equation~\eqref{theorm-eq-1} using our method, which converged in 0.3 seconds, are illustrated in Figure~\ref{fig:DIST_DIF_COST}.

\begin{figure*}
\vspace{-6mm}
    \begin{subfigure}[t]{0.35\textwidth}
        \centering
\includegraphics[width=\textwidth]{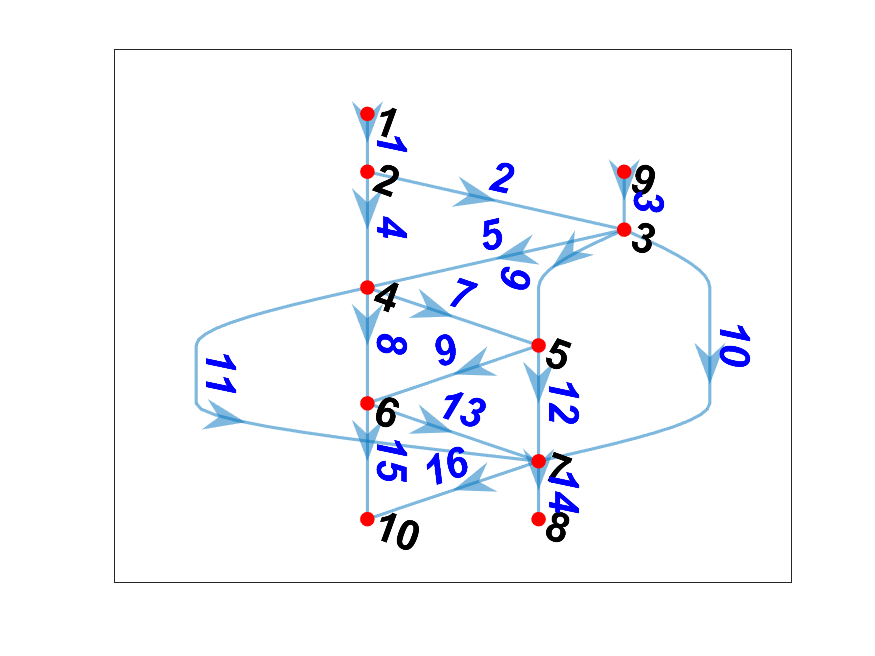}
    \caption{Full road graph with edge length ($s_k$).}
    \label{fig:FULL_GRAPH_EM}
    \end{subfigure}
    ~
    \begin{subfigure}[t]{0.35\textwidth}
        \centering
\includegraphics[width=\textwidth]{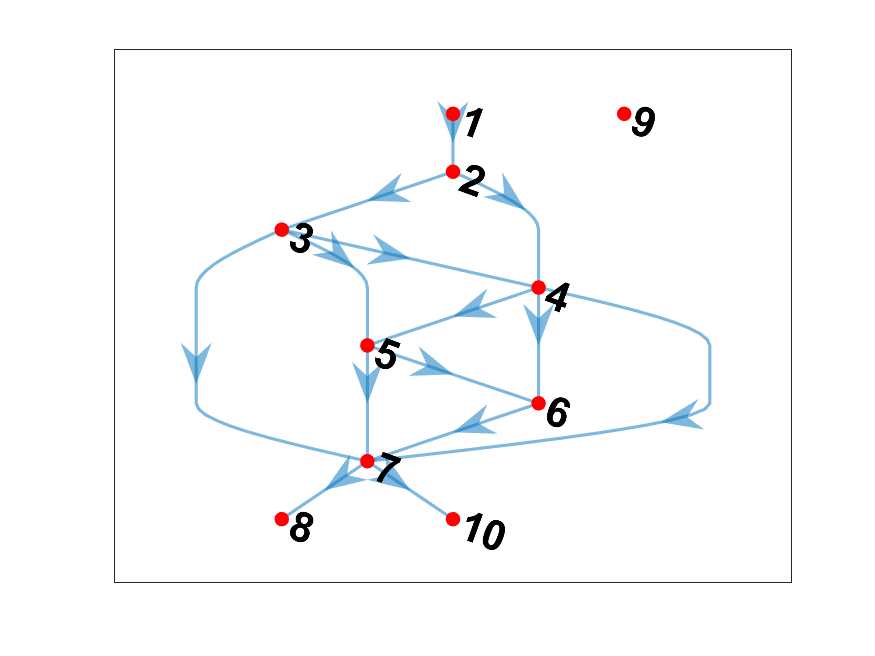}
        \caption{Car flow sub-graph.}
        \label{fig:Pop1_EM}
    \end{subfigure}%
    ~
    \begin{subfigure}[t]{0.35\textwidth}
        \centering
\includegraphics[width=\textwidth]{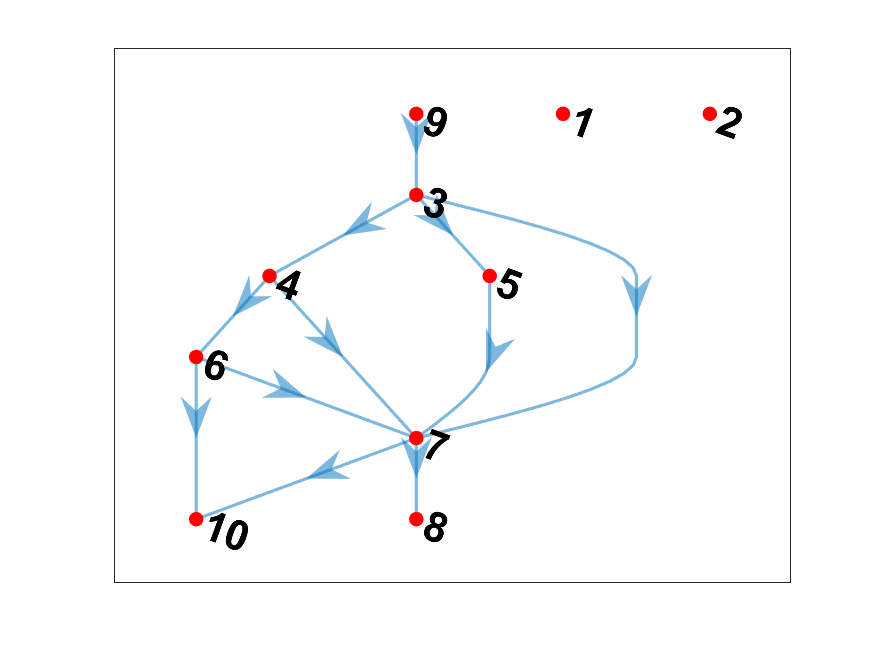}
        \caption{Truck flow sub-graph.}
        \label{fig:Pop2_EM}
    \end{subfigure}    
    \caption{Individual Population sub-graphs for flow Emissions.}
    \vspace{1mm}
\end{figure*}

\subsection{Reducing Emissions Cost with Traffic Management}
\begin{figure}[h]
\vspace{-5mm}
\centering
    \begin{subfigure}[t]
    {0.4\textwidth}
        \centering
\includegraphics[width=\textwidth]{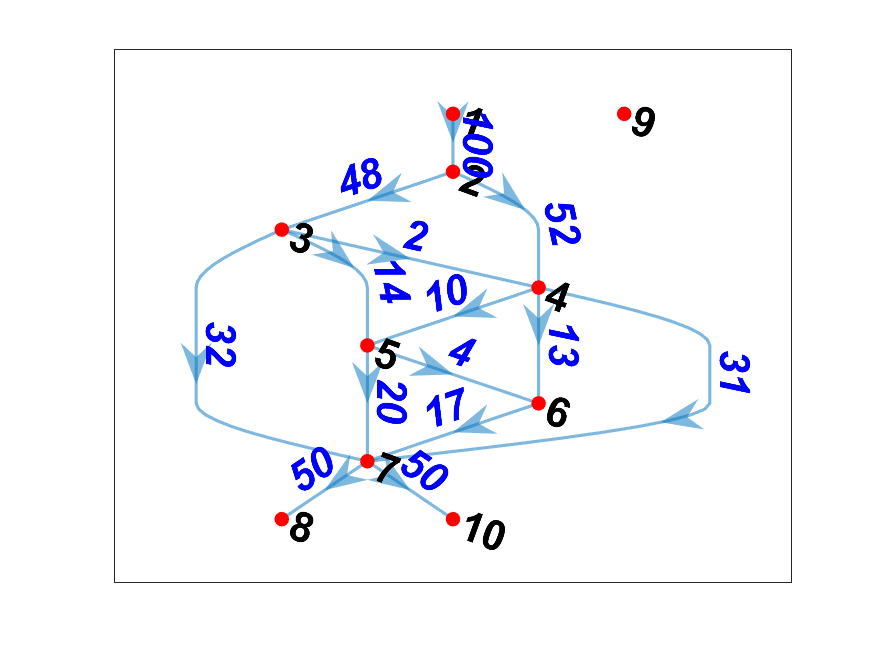}
        \caption{Minimum emission car flow sub-graph.}
        \label{fig:Pop1_em_res}
    \end{subfigure}%
    \\
    \begin{subfigure}[t]{0.4\textwidth}
        \centering
\includegraphics[width=\textwidth]{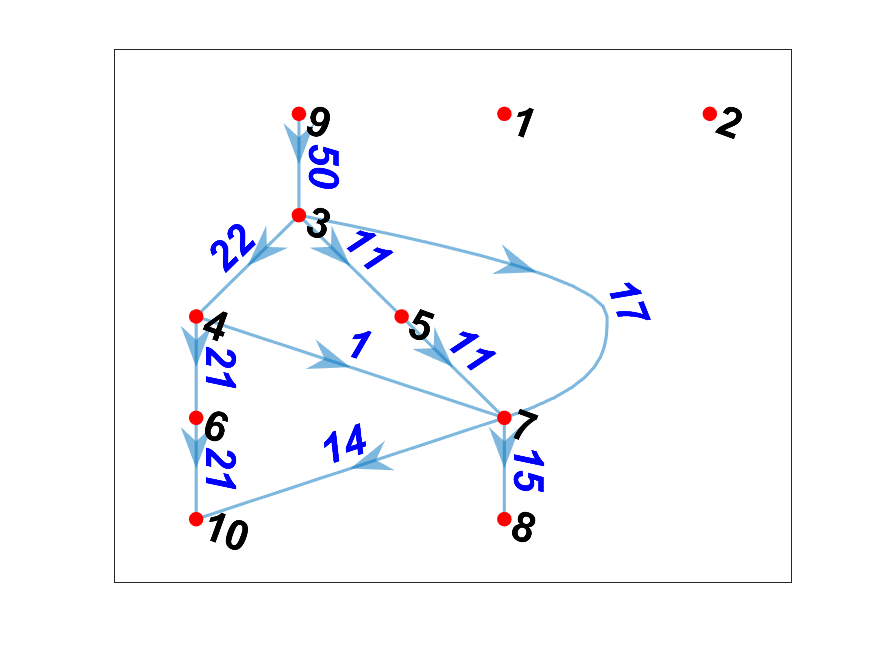}
        \caption{Minimum emission truck flow sub-graph.}
        \label{fig:Pop2_em_res}
    \end{subfigure}    
    \caption{Minimum emission individual populations.}
\label{fig:DIST_DIF_COST_EM}
\end{figure}

One major problem in dense urban environments arises from vehicle emissions, which impact both public health and fuel consumption costs. In \cite{Emissions}, models for fuel consumption (FC) and emissions—specifically Hydrocarbons (HC), Nitrous Oxides (NOx), Carbon Monoxide (CO), and Carbon Dioxide (CO2)—are discussed. Particularly, the emissions are calculated using the average speed of the vehicles on every edge, which is affected by the total flow in the respective edges. The average speed described by the flow on the edge:
\begin{equation*}
       v_k(\jmath_k) = \tfrac{s_k}{t_k }\left(1+\alpha_k \left(\tfrac{\jmath_k}{\kappa_k}\right)^{\beta_k}\right)^{-1},
\end{equation*}
where \(s_k>0\) is the length of edge \(k\), \(t_k>0\) is the free-flow travel time, \(\alpha_k>0\) and \(\beta_k>0\) are parameters that quantify congestion effects, \(\kappa_k>0\) is the practical capacity of edge \(k\), and $\jmath_k$ represents the total flow of all population inside the edge. On each edge $k$, and for each type of vehicle $r$, the emission rates are calculated for each type of emission $j$, \(j \in E = \{FC, HC, NOx, CO, CO2\}\). As such, the emission rate $(\text{g}/\text{km})$ is dependent on the average speed of the flow on each respective edge by the following equation:
\begin{equation*}
    e_{\{k,j\}}^r(\jmath_k) = \tfrac{a^r_j}{v_k(\jmath_k)} + b^r_j.
\end{equation*}
Here, \(a^r_j\), and \(b^r_j\) are parameters calibrated for FC, HC, NOx, CO, and CO2 for each vehicle type, with \(a^r_j>0\).
Finally, the emission cost for edge \(k\) and population \(r\) is 
\begin{equation*}
    c_k^r(J) = \tfrac{s_k \sum_{j \in E} w_j\, e_{\{k,j\}}^r(\jmath_k)}{2} + \tfrac{\jmath^r_k}{2},
\end{equation*}
where \(w_j\) denotes the cost of emission \(j\) in \(\$/\mathrm{kg}\). The values of \(a^r_j\) and \(b^r_j\) are given in Table~\ref{tab:param_em}. In this setup, each population minimizes its own emissions (and thereby congestion) and aims to reduce its concentration by spreading evenly across the network. This helps prevent high emission levels in localized areas. For the simulations, we consider two classes of vehicles: cars and trucks. Trucks are assumed to have twice the impact on congestion and emit three times as much pollution compared to cars, i.e., \( e^2_{\{k,j\}} = 3e^1_{\{k,j\}} \).
\begin{table}[]
\vspace{-5mm}
\begin{tabular}{|c|c|c|c|c|c|c|c|}
\hline Objective ( $\mathrm{g} / \mathrm{km}$ ) & a & b & w (\$/kg) \\
\hline Fuel consumption (FC) & $1.56 \times 10^{3}$ & $3.54 \times 10^{1}$ & 1.0321 \\
\hline Hydrocarbons (HC) & $1.08 \times 10^{1}$ & $-7.11 \times 10^{-3}$ & 12.91 \\
\hline Nitrous Oxides
(NOx) & $2.00 \times 10^{0}$ & $-4.49 \times 10^{-2}$ & 14.54 \\
\hline Carbon Monoxide
(CO) & $8.08 \times 10^{1}$ & $1.16 \times 10^{0}$ & 0.37 \\
\hline Carbon Dioxide
(CO2) & $4.78 \times 10^{3}$ & $1.11 \times 10^{2}$ & 0.02 \\
\hline 
\end{tabular}
\caption{Parameters for each emission type for standard cars under optimal speed.}
\label{tab:param_em}
\vspace{-7mm}
\end{table}
In the simulation, we set $t_k=\frac{50}{s_k}$, where $50 \mathrm{km}/\mathrm{hr}$ is the free flow speed for all edges. Moreover, we set  $\beta_k=3$, $\alpha_k=5$, and $\kappa_k=50$ for all $k$s. The new graph and the respective edge lengths are shown in \ref{fig:FULL_GRAPH_EM}. The subsequent subgraphs are shown in Figures \ref{fig:Pop1_EM} and \ref{fig:Pop2_EM}.
It can be seen that there are unique edges for each population. Edge $ (6,10) $ is a truck road, while edges $ (4,5) $ and $ (5,6) $ are only car roads. This is possible since our reformulation of the congestion problems allows us to define unique subgraphs for each population. The simulation was done using the same method as the previous scenarios, and the result converged within less than $1$ second.
The resulting flows can be seen in Figures \ref{fig:DIST_DIF_COST_EM}. The total emission cost amounted to $\$4751$, which is a drastic change from any other random distribution of populations along the network, which can amount to the orders of $\$10^6$. Moreover, as one can see from Figures \ref{fig:DIST_DIF_COST_EM}, the trucks and cars are evenly spread out across the network (e.g. car flows between in $(7,8)$ and $(7,10)$ in Figure \ref{fig:Pop1_em_res}), minimizing the concentration of emissions in concentrated areas and allowing the equal spread of emissions. 

\section{Conclusion}\label{sec:Conc_Future_work}
In this paper, we develop an analysis and solution for the Multi-Population Wardrop Problem. We formulate the single and multi-population setup by transforming the search problem into a variational inequality with equality constraints. Furthermore, we establish conditions for the existence and uniqueness of the Multi-Population Wardrop Equilibrium and analyze its relation to the Nash Equilibrium of the populations. We then reformulate the search problem as a distributed optimization problem, for which we introduce the Hessian Riemannian flow approach. We prove that solving the associated ordinary differential equation (ODE) leads to the unique Multi-Population Wardrop Equilibrium Problem solution under appropriate conditions. Finally, we apply this solution to three scenarios, highlighting the importance and efficiency of our proposed method.




\bibliographystyle{ieeetr}

\bibliography{bibtex/bib/IEEEabrv}

\begin{thebibliography}{10}

\bibitem{Wardrop_Equilibria}
J.~R. Correa and N.~E. Stier-Moses, {\em Wardrop Equilibria}.
\newblock John Wiley and Sons, Ltd, 2011.

\bibitem{Uniqueness_general}
A.~Orda, R.~Rom, and N.~Shimkin, ``Competitive routing in multi-user communication networks,'' in {\em IEEE INFOCOM '93 The Conference on Computer Communications, Proceedings}, vol.~3, pp.~964--971, 1993.

\bibitem{Con_Multi}
M.~Gairing and M.~Klimm, ``Congestion games with player-specific costs revisited,'' in {\em Algorithmic Game Theory} (B.~V{\"o}cking, ed.), (Berlin, Heidelberg), pp.~98--109, Springer Berlin Heidelberg, 2013.

\bibitem{Wardrop_Nash}
A.~Haurie and P.~Marcotte, ``On the relationship between nash—cournot and wardrop equilibria,'' {\em Networks}, vol.~15, no.~3, pp.~295--308, 1985.

\bibitem{Topological_Nash}
O.~Richman and N.~Shimkin, ``Topological uniqueness of the nash equilibrium for selfish routing with atomic users,'' {\em Mathematics of Operations Research}, vol.~32, no.~1, pp.~215--232, 2007.

\bibitem{atomic_not_unique}
U.~Bhaskar, L.~Fleischer, D.~Hoy, and C.-C. Huang, {\em Equilibria of Atomic Flow Games are not Unique}, pp.~748--757.

\bibitem{n_games}
J.~B. Rosen, ``Existence and uniqueness of equilibrium points for concave n-person games,'' {\em Econometrica}, vol.~33, no.~3, pp.~520--534, 1965.

\bibitem{Polynomial_cost}
E.~Altman, T.~Jimenez, T.~Basar, and N.~Shimkin, ``Competitive routing in networks with polynomial cost,'' in {\em Proceedings IEEE INFOCOM 2000. Conference on Computer Communications. Nineteenth Annual Joint Conference of the IEEE Computer and Communications Societies (Cat. No.00CH37064)}, vol.~3, pp.~1586--1593, 2000.

\bibitem{Survey}
N.~Shimkin, ``A survey of uniqueness results for selfish routing,'' in {\em Network Control and Optimization} (T.~Chahed and B.~Tuffin, eds.), (Berlin, Heidelberg), pp.~33--42, Springer Berlin Heidelberg, 2007.

\bibitem{Beckman}
C.~B.~M. Martin~Beckmann and C.~B. Winsten, ``Studies in the economics of transportation,'' {\em The Economic Journal}, vol.~67, pp.~116--118, 03 1957.

\bibitem{TRUCKS_AND_CARS}
H.~S. Mahmassani and K.~C. Mouskos, ``{Some numerical results on the diagonalization algorithm for network assignment with asymmetric interactions between cars and trucks},'' {\em Transportation Research Part B: Methodological}, vol.~22, pp.~275--290, August 1988.

\bibitem{Lemke-like}
F.~Meunier and T.~Pradeau, ``A lemke-like algorithm for the multiclass network equilibrium problem,'' in {\em Web and Internet Economics} (Y.~Chen and N.~Immorlica, eds.), (Berlin, Heidelberg), pp.~363--376, Springer Berlin Heidelberg, 2013.

\bibitem{Lemke-line-2}
F.~Meunier and T.~Pradeau, ``Computing solutions of the multiclass network equilibrium problem with affine cost functions,'' {\em Annals of Operations Research}, vol.~274, pp.~447--469, Mar 2019.

\bibitem{Hessian}
F.~Alvarez, J.~Bolte, and O.~Brahic, ``Hessian riemannian gradient flows in convex programming,'' {\em SIAM Journal on Control and Optimization}, vol.~43, no.~2, pp.~477--501, 2004.

\bibitem{Hessian_MONOTONE}
{Gomes, Diogo A.} and {Yang, Xianjin}, ``The hessian riemannian flow and newton’s method for effective hamiltonians and mather measures,'' {\em ESAIM: M2AN}, vol.~54, no.~6, pp.~1883--1915, 2020.

\bibitem{Var_prob}
P.~T. Harker and J.-S. Pang, ``Finite-dimensional variational inequality and nonlinear complementarity problems: A survey of theory, algorithms and applications,'' {\em Mathematical Programming}, vol.~48, no.~1, pp.~161--220, 1990.

\bibitem{Wardrop1}
F.~A. Saleh, T.~Bakaryan, D.~Gomes, and R.~de~Lima~Ribeiro, ``First-order mean-field games on networks and wardrop equilibrium,'' {\em Portugaliae Mathematica}, vol.~81, no.~3/4, pp.~201--246, 2024.

\bibitem{khalil2002nonlinear}
H.~K. Khalil, {\em Nonlinear Systems}.
\newblock Upper Saddle River, NJ: Prentice Hall, 3rd~ed., 2002.

\bibitem{Emissions}
A.~R. James~Tidswell, ``Modelling traffic assignment objectives with emission cost functions,'' (Auckland, New Zealand), Australasian Transport Research Forum, 2013.

\end{thebibliography}

\end{document}